\documentclass[12pt,a4]{article}
\input{epsf.tex}
\begin{document}
\begin{center}
{\bf {THE 
PROBABLE LEVEL DENSITIES AND RADIATIVE STRENGTH FUNCTIONS OF
DIPOLE GAMMA-TRANSITIONS IN $^{57}$Fe COMPOUND NUCLEUS}}
\end{center}
\begin{center}{\bf A.M. Sukhovoj, V.A. Khitrov, Li Chol}\\{\it 
FLNP, Joint Institute for Nuclear Research, Dubna, Russia}\\{\bf
Pham Dinh Khang}\\{ \it National University of Hanoi}\\{\bf
Vuong Huu Tan, Nguyen Xuan Hai}\\{\it
Vietnam Atomic Energy Commission}\\
\end{center}
%\parbox{155mm}
\begin{abstract}
From the  published results of experimental research of $^{56}$Fe(n,$2\gamma$)
reaction carried out in 
Budapest, the values of probable densities of cascade intermediate levels with 1/2, 3/2 spin 
and radiative  strength  functions of cascade E1 and M1 transitions in 
$^{57}$Fe compound nucleus have 
been determined. These results correspond to analogous data for other nuclei
studied by us and 
contradict to predictions of conventional models.
\end{abstract}
\section*{Introduction} \hspace*{14pt}
The modern experiment techniques permit one to measure parameters of  most nuclear 
reactions with error about 10\% or less. Meanwhile, specific characteristics of nucleus
which are 
important for  theory and practice such as level density and radiative  strength functions of 
cascade transitions are derived from experiment with less accuracy. The principal cause
of this
is insufficient volume of experimental information used in fitting process or
incorrectness of 
its algorithms. 
The level density and strength functions of cascade gamma-transitions below neutron binding 
energy $B_n$ are typical examples of problems to elucidate experiment results 
with high reliability levels.
This is certainly when:

1. Setting up the best method to find unknown quantities from experimental spectra if they 
cannot be determined directly.

2. Collecting the experimental data as much as possible to guarantee the singularity of the 
desired parameters and the ability to use these quantities for complementary independent 
revision.

3. Evaluating all the sources of systematic error of achievable values.

Actually, the level density $\rho$ was determined now from experimental data on:

	(a) the nucleon evaporation spectra (see, for instance, [1]);
	
	(b) the intensities of two-step gamma-cascades [2]
\begin{equation}
I_{\gamma\gamma}=\sum_{\lambda ,f}\sum_{l}\frac{\Gamma_{\lambda l}}
 {\Gamma_{\lambda}}\frac{\Gamma_{lf}}{\Gamma_l}=\sum_{\lambda ,f} 
\frac{\Gamma_{\lambda l}}{<\Gamma_{\lambda l}> m_{\lambda l}} n_{\lambda l}
\frac{\Gamma_{lf}}{<\Gamma_{lf}> m_{lf}},
\end{equation}
between compound state (neutron resonance) and  a group of low-lying  levels of
studied nucleus 
determined according to algorithm [3] for all possible intervals $\Delta E$ of energy $E_1$
of cascade 
primary gamma-transitions;

	(c) the gamma spectra depopulating  levels with the excitation energy $E_{ex}$
in  (d,p) [4] and $(^3$He,$\alpha$) [5] nuclear reactions;.

Within methods [2] and [5],  $\rho$ is determined simultaneously with strength functions of
cascade gamma-transitions:
       \begin{equation} k=f/A^{2/3}=\Gamma_{\lambda l}/(E_{\gamma}^3\times A^{2/3}
\times D_{\lambda}).
\end{equation} 
Here $E_{\gamma}$ is the gamma-transition energy and  $D_{\lambda}$ is the  distance between
initial levels $\lambda$.
Such form of $k$ provides minimal dependence of this parameter on nuclear mass and
allows direct 
comparison of these parameters in nucleus with different masses.
It is assumed that the partial radiative 
width  $\Gamma_{\lambda l}$ of gamma transition between  $\lambda$  and  $l$  levels is the
random average value, and the 
width ratio for different multipolarity transitions can be determined  experimentally. It
is evident that the criterion of the necessary confidence for the determination of $k$
and  $\rho$
does not 
cause the difference among the results of various experiments in principle.
In other words, once there are errors, explanation for these is obligatory.
The results of $^{57}$Fe compound nucleus [5,6] are 
ideal examples in this situation.

\section{The basic principles to determine $\rho$ and $k$ from intensities of two-step
cascades}
\hspace*{14pt}

Equation (1) contains more unknown parameters than measured the number of intervals in it.
Nevertheless, this 
type of relation between  $\rho$  and $k$ with $I_{\gamma\gamma}$ provides limitation of the
region of their probable 
values (the simplest example for such type of function  is giperellipsoid $\sum (ax_i)^2$=const 
with arbitrary variable number $i$). The  interval of $\rho$ and $k$
possible values is rather narrow only if experimental spectra are decomposed in sole 
primary and sole secondary components according to [3], and ratio $\Gamma_{\lambda l}$,
$\Gamma_{lf}$  of primary and
secondary transitions widths with equal energy and a multipolarity is set in some manner.
Although the first condition is not satisfied for experimental data from [6],
the use of algorithm [2] and data from files of ENDS [7] and EGAF
[8] permits one to find narrow enough intervals of  $\rho$ and $k$ values
(with the accuracy ($\chi^2/f<1$) which
provide reproduction of $I_{\gamma\gamma}$ in all 27 energy intervals of
cascade gamma-transitions. In order to 
get this results, the current $k$ values of iterative process [2] have to be replaced by the
values which are set by experimental intensities of primary transitions 7.645, 7.279, 6.380,
6.018, 5.915, 4.217, 3.436 and 3.266 MeV.
As usually, in iterative process we used values of neutron resonances,
low-lying  levels densities and total radiative width of s-resonances [9].
Level density  of $^{57}$Fe at the  thermal neutron capture
is very small, therefore the results of the method  used to 
determine $\rho$  and $k$ are sensitive enough for parameters and decay types of low-lying  levels involved in
calculation of cascade intensity. Corresponding data on 10 levels with $E_{f} \leq 2.113$ MeV
with spins $J^\pi=1/2^{-}-7/2^{-}$ were taken from ENDS-file.
Radical discrepancy between these results and the $\rho$ and $k$ values from [5,6]
has the simplest 
explanation: reliable data can be extracted from the experiment only under condition of 
execution of basic principles of mathematics and mathematical statistics.
First of all, should 
be taken into account  specific of errors transfer between the measured spectrum and parameters 
derived from it. 
From eq. (1), it follows, that in the first approach, the error of cascade intensity
 \begin{equation}
 \delta I_{\gamma\gamma}(E_1)=\sum (dI/d{\rho}\times \delta \rho+dI/dk \times \delta k)
\end{equation}  
is the incoherent nonlinear sum of errors  $\delta \rho$ and $\delta k$ from {\bf whole}
interval of possible energies of
the cascade intermediate levels. This means, in particular, that the coincidence of the 
experimental and calculated  cascade intensities for given $\rho$ and $k$ 
in individual intervals of nuclear excitation cannot be interpreted (as it was done in [6])
as the proof of their equality to unknown values. 
This fact is illustrated in Fig. 1. Here experimental intensities were obtained as 
superposition of graphic data in figs. 2 and 3 from [6]. They are compared with:
 
   (a) the calculation which uses the results [5] of determination of $\rho$ and $k$ 
   from the data on reaction $^{57}$Fe($^3$He,$^3$He,$\gamma)^{57}$Fe and
   
   (b) the typical results of calculation of $I_{\gamma\gamma}$
   for both obtained like in [2] random $\rho$ and $k$ values
from the interval of their possible variation, and variant in which level density is fixed
according 
to data [5] and only strength functions are varied (see Figs. 2 and 3).

	In case of analysis of the experimental spectrum without its decomposition [3] into 
components corresponding to solely primary and solely secondary transitions, the total error of
intensity includes in additional error related with registration of cascades with
secondary transitions from the same energy interval: $\delta I_{\gamma\gamma}^{exp}(E_1) =
\delta I_{\gamma\gamma}^{prim}+\delta I_{\gamma\gamma}^{sec}$. This leads to 
additional widening of interval of the $\rho$ and $k$ which reproduce experimental
spectra due to false 
solutions. For example, fixation of level density in method [2] according to the data [5]
results in 
the sum strength functions which, in the regions  $\sim 3.2$ and $\sim 4.4$
MeV, are more than two times 
larger than the maximum experimental value of $k$.
 This means that the level density in $^{57}$Fe listed 
in [5] contains some systematic error.

\subsection{ The spectrum of possible functions $\rho$, $k$ and their most probable values
} \hspace*{14pt}
Values $\rho$ and $k$ determined within method [2] contain statistic errors.
Apart from above
classical errors, when calculating the unknown values, the spectra of $\rho$ and $k$ 
are stretched by the degradation of (1) equation system. 
This equation system is nonlinear and the correlation
coefficients absolute values  between their quantities are close to unit
for a given $\chi^2$ only for some part of $\rho$ and $k$ values.
The remained cases show correlation
for other pairs of parameters of (1) system. 

It makes impossible situation for degrade linear equations system  where interval of
solution equals infinity.
So, in the case considered here, the interval of all possible $\rho$ and $k$
values is always finite.
The sign of $\delta I_{\gamma\gamma}(E_1)$ is different with equal probability 
at least around its zero value) and
probability of random localization of iterative process [2]  in vicinity of point,
for example, 
$\sum (dI/d{\rho}\times \delta \rho)=-\sum (dI/dk \times \delta k)$ 
is inverse to the value $\sum (|dI/d{\rho}\times \delta \rho|+|dI/dk \times \delta k |)$.

Therefore, one can make conclusion about asymptotically  convergence  of found
solutions to most
probable values, i.e., the average values of ensemble of N realizations of iterative process 
[2] with corresponding  standard deviation.
Therefore, the $\rho$ and $k$ functions estimated in this way differ significantly the from
data [5].
Level density (Fig. 2) has clearly   expressed ``step-like" structure as, in practice,
in other nuclei studied by means of method [2].
The variations of the sum of radiative strength functions of dipole
transitions (Fig. 3) directly testify to its dependence on the structure of cascade
 intermediate levels in region $B_n$
(if only intensities of cascades  [6] with transition energy less than 2.0-2.5 MeV  do
not have large systematic  errors).
First of all, it should be taken into account that the structure in
cascade intensities presented in Fig. 1 in interval $2.5 \leq E_\gamma \leq 5.5$ 
MeV  cannot be reproduced in
calculation using level densities and strength functions from [5,6].
In correspondence with [3], we could come to a conclusion that data on level
densities and strength functions listed in [5,6] cannot reproduce total intensity
of two-step cascades even in
principal.
In general case, two explanations for the disagreement between results of methods [2] 
and [5,6] could be accepted:

   (a)  the presence of serious systematic errors in extraction of $\rho$ and $k$
   from $I_{\gamma\gamma}$ and from the total
gamma-spectra corresponding to decay of $^{57}$Fe levels excited in ($^3$He,$^3$He$,\gamma$)
reaction;

   (b) the difference of $\rho$ and $k$ values reproducing the  spectra of these
two experiments is can result by details 
of excited levels wave functions
structures. 
Of course, the second possibility is to be taken into account only in case of
small 
experimental systematic errors. 

\section{Main sources of systematic errors in various methods to determine $\rho$ and $k$ 
values and any
ways of their reduction}\hspace*{14pt}
So, principal differences in level densities [1,5] and 
[2] leads to necessity to examine data analysis algorithm and to
estimate the confident 
level of results.

In methods [1,5], common problem at determination of $\rho$ is the relationship between number
$m=\rho \times \Delta E$ of levels excited in a given energy interval $\Delta E$
and probability  $T$ (width $\Gamma$) of measurable reaction product emission.
 Besides, the corresponding spectra can be reproduced by infinite 
number of functional dependencies  $\rho=f(E_{ex})$ and  $T=\phi(E_{ex})$,
the intervals  of possible values of
level density and emission probability of reaction product $T$ can vary from  $-\infty$
 to $+\infty$. It means
that the type of spectra measured with methods like [1,5] always provides the result
with lower 
confident level as compared with  method [2].
 The intensity measured within method [2], to the first
approach, is inverse to level density and proportional to $dk/dE_\gamma$.
Therefore, in all circumstances, 
this provides results with better confidence as compared with methods [1,5].
Again, it should be
taken into account that, analysis of spectra containing [2] only from events of full energy
registration of two-step cascades with fixed $J^{\pi}$ of their initial and final
levels always gives considerably better accuracy than analyses the total gamma-ray
spectra measured with
scintillation detector, even though owing to supplementary systematic errors
caused by  subtracting [13]
Compton background, and due to higher stability of system with Ge detectors. 

In addition, in case [1], determinetion of experimental value $\rho$,
it required the use of theoretical
value $T$ of penetration of nucleus surface for evaporated  nucleon (or light nucleus).
This value
should be calculated in frame of nuclear models  with error being not larger than some
tens percents at possible high-frequent
sign-changeable variations of $T$ relatively to average value. 
The existence of $T$ gradient directly
follows from results of modern method [14] to extract $\rho$ and $k$ 
from  intensities of two-step
cascades due to high precision of corresponding experiment.  As a result, we cannot evaluate
confidence of data like [1] objectively.
Therefore, it is impossible to use these data to confirm [5] results.

First of all, there is no objective evidence to reject in methods [1,5]
the possibility for compensation of level density
increasing (decreasing)  in some excitation energy interval by the decreasing
(increasing) of neutron or gamma-quantum emission probability in evaporation
or primary gamma-transitions
spectrum at the decay or excitation of levels close to $E_{ex}$. This
compensation is directly observed [14] for $\rho$ and $k$ extracted from
$I_{\gamma\gamma}$ data. The optical potentials
used in method [1], most probably, cannot provide the calculation $T$ with required
details and errors of $\rho$ at the level achieved in framework of method [14].
The evaluation [15,16] of statistical error and degree of its influence 
on determined in  [2] values of $\rho$ and $k$ showed that method [2]
reproduces the peculiarities of above parameters for all possible ordinary
experimental systematic errors. 

The situation in data analysis from $(^3$He,$\alpha\gamma)$ reaction  is rather
worse. The simple analysis [17] of error transfer from primary transition 
spectra  determination 
to the fit of this spectra by ''Oslo method'' [18] with some $\rho$ and $k$ values
demonstrates that systematic errors of the desired parameters 
increase at least by a factor of several tens at every step.
 At the first step of [17], it is caused by the fact that
the intensities of primary transition spectrum are much less than intensities of total
gamma-spectra (Fig. 4).
That is firstly related to the low energies of primary gamma-transitions. 
It is easy to get the lower estimation of sum of statistical errors of the primary
gamma-transition spectra from the  narrow energy interval in the vicinity of $E_{ex}$.
This requires one to 
normalize (per 1 decay)   all full energy spectra of [17] according to equation:	
\begin{equation}
 \sum I_\gamma \times E_\gamma=E_{ex}.
\end{equation}
The deviation of area of the primary transition   spectrum [17] from unity is the part of
sum of its unknown errors.
The maximum of this sum deviation from unit should be very small in order to get
acceptable systematic errors of $\rho$ and $k$.

At the second step, because of the strong correlation [18] between $\rho$ and $k$,
the increase in its relative errors is practically of the same scale. 
For example, in order to get error of $\rho$ and $k$ 
about $\sim 50\%$, it is necessary [19] to determine (and normalize to 1) total
gamma-spectra
at arbitrary $E_{ex}$ with systematic error less than $\sim 0.1-0.3\%$.
It is possible to reject or reduce the systematic errors by using modified variant of
method [17] to get spectra of primary transitions according to eq. (4) and to determine
the best values $\rho$ and $k$. For example, to use the Gauss method and
to correct degrade matrix, this
allows [20] to realize method of search for the likelihood function maximum
within two main principles of mathematical statistics:

(a) to determine supplemental vector of corrections at each iteration
by the Jacobean matrix without any assumption;

b) to study explicit form of maximum of $\rho$ and $k$ likelihood function
and to select false maxima by simple variation of initial $\rho$ and $k$. 
The method [18] does not fully satisfy two above conditions.

The situation with the confidence level of $\rho$ and $k$ obtained within method
[2] is much better. 
Of course, level density and strength functions derived from experimental
$I_{\gamma\gamma}$
data  include both 
ordinary and specific systematic errors.
Possible ordinary errors of $\rho$ and $k$ can be easily determined by estimating the
propagation error of experimental cascade transition intensity to values
of $\rho$ and $k$.
For example,
comparison between the intensities of the cascade primary transitions from [8] and
[21] used for
normalization $I_{\gamma\gamma}=F(E_1)$ and the use of results [22] of
extrapolation to zero threshold for
distribution of random cascade intensities for intermediate levels energies
$E_{ex} < 0.5$ MeV permits precise
determination systematic errors for both  amplitude and form of
$F(E_1)$ function. The variation of initial data [2] in limits of
their expected errors allows to estimate [20]
ordinary systematic errors of $\rho$ and $k$ at presence of nonlinear and ambiguous
relation between 
$\delta F(E_1)$, $\delta \rho$ and $\delta k$.
 As it was shown in [15], these $\delta \rho$ and $\delta k$ values obtained
in this manner cannot
explain ``step-like" structure of  $\rho$ and $k$ [2] by discrepancy between
these parameters and conventional
notions about smooth energy dependence of level density.

\section{ The ability to reduce special systematic errors in determination
of $\rho$ É $k$}\hspace*{14pt} 
The left part of eq. (1) is determined by three unknown functions: the total density
of the
cascade intermediate levels in a given energy interval, sums of radiative strength
functions of
primary and secondary dipole transitions. Strong anti-correlation of these parameters
causes the
phenomenon that total level density with different parity and spin (this interval
is determined by
multipolarity selection rule), and total strength functions of E1 and M1 transitions
exactly ($\chi^2/f <<1$) 
reproducing $I_{\gamma\gamma}$ vary in narrow enough interval.
Eventually, this interval for level density
with  $\pi=+$, $\pi=-$, $k(E1)$ and $k(M1)$ separately is always wider than
for their sums. This
statement is true if the primary and secondary transitions width ratio is set on the
basis of some information. Without the above information, the only way to determine
$\rho$ and $k$ in
[2] method is the use the assumption of equal energy dependencies of k for primary and
secondary transitions. (Modern nuclear models clearly show incorrectness of this
hypothesis at excitation energy of about several MeV [23] and, moreover,
require [24] detailed accounting for nuclear structure
at least  up to 20 MeV excitation energy).

The partial compensation of incorrectness of assumption $k^{prim}/k^{sec}=const$ used  in [2] is
provided by sign-variable deviation of  $k$ for secondary transitions at various energy with respect 
to that for primary transitions. It means that the relative variation of
total radiative width $\Gamma_l$ of
cascade intermediate level in method [2] really is considerably less than the relative
change the energy dependence of k for secondary transition with respect to that
for primary transitions. This 
is observed in all nuclei where experimental data permit one to use [14] method to evaluate the
general trend of $k(E_\gamma, E_{ex})$ function.
Furthermore, level density in all nuclei  obtained within method
[14]  has more  serious discrepancy with the predictions according to [12] than it was earlier
established in [2].
The effect of change in  ($k^{prim}/k^{sec} \neq const$) for different energies of decaying level on 
parameters to be determined according to [5] can be stronger because of the used in [18]
averaging of strength functions for different  energies of decaying levels.

\section{Evaluation of confidence level for $\rho$ and $k$ parameters
and possibility to improve it}\hspace*{14pt}
The experimental data on $^{57}$Fe include gamma-ray spectrum following thermal neutron
radiative capture. These data are suitable, at least, to reveal the difference between the
experimental and estimated $\rho$ and $k$ values. Comparison between the experimental and calculated 
spectra for different level densities and radiative strength functions is done in Fig. 5. It confirms
conclusion that models like [10-12] cannot be used for precise description of the compound nucleus cascade
gamma-decay.
It also testifies to the necessity of further analysis of the characters of this process, reduction 
and more detailed accounting for influence of all the sources of systematic errors on the
parameters under search. First of all, to reduce error of $I_{\gamma\gamma}=F(E_1)$
and  determine in details energy
dependence of function $k=\phi(E_\gamma, E_{ex})$ directly from the experiment. 
This may be done after
determination of intensities of two-step cascades to final levels with  $E_{ex}>0.5B_n$
in new experiment with higher quality.
Otherwise, one can also to reanalyze experimental data [6]. 

Using numerical method [26] to improve energy resolution together with method [27] to
determine quantum ordering in resolved
{in form of peak pairs) cascades and known decay scheme [7], one can determine
function (1)  with small enough systematic error.
From the population of intermediate levels determined according to [14] from intensities of
resolved cascades, one can get notion of general tendencies of change in strength functions as
changing energy and involve these data in analysis [2]. This will allow use to reject
a part of false
$\rho$ and $k$ values obtained in analysis of row  cascade intensities and,
to the first approach,
 to take
into account the change in ratio of intensities of cascade secondary transitions  
to final levels
with different  excitation energy and deeply different  wave function structures.
\section{Conclusions} \hspace*{14pt}
The analysis (according to the basic principles of mathematics) of the two-step cascade
intensities following 
thermal neutron capture in $^{56}$Fe  confirms conclusion obtained earlier for other nuclei that 
correspondence between the experimental and calculated [10-12] level densities and radiative 
strength functions within experimental errors cannot be achieved. The obtained level density 
corresponds to general trend observed in [2,14] for other nuclei. According to theoretical notions 
[28], this phenomenon is caused by breaking in this nucleus of at least two Cooper pairs of 
nucleons that results also in change in ratio of portions of quasi-particle and vibrational 
excitations. This phenomenon is not taken into account in the models like [10,12].
Significant increase in strength functions for $E_1<2$ MeV  completely corresponds [14] to 
localization of step-like structures in energy dependence of level density for secondary 
transitions of cascades. In totality, the $\rho$ and $k$ values derived from experimental data [6] testify to 
the effect of nuclear structure on these quantities in wide energy ranges and some variation of 
this effect for different nuclei.  Perhaps, from position of investigated nucleus with
respect to full nucleon shells number.
\newpage
{\bf References}\\\\
1. Zhuravlev B.V., Lychagin A.A., Titarenko N.N., Trykova V.I.,\\
\hspace*{14pt} Interaction of Neutrons with Nuclei: Proc. of XII 
International Seminar, Dubna,\\\hspace*{14pt} May 2004,
E3-2004-169, Dubna, 2004, p. 110.\\
2. Vasilieva E.V., Sukhovoj A.M., Khitrov V.A.,
 Phys. At. Nucl. 2001, V. 64, $N^{o}2$, 153.
\\ \hspace*{14pt}  Khitrov V. A., Sukhovoj A. M.,
 INDC(CCP)-435, Vienna, 2002, p. 21.\\
\hspace*{14pt} http://arXiv.org/abs/nucl-ex/0110017.\\
3. Boneva S. T., Khitrov V.A., Sukhovoj A.M.,
 Nucl. Phys. 1995, {\bf A589} 293.\\
4. Bartholomew G.A. et al.,
Advances in nuclear physics, 1973, {\bf 7} 229.\\
5. Schiller A. et all, Phys.\ Rev.,\ C, 2003, {\bf 68(5)} 054326-1.\\
6. Voinov A. et all, Phys.\ Rev. Let.,\ 2004, {\bf 93(14)} 142504-1.\\
7. http://www.nndc.bnl.gov/nndc/ensdf.\\
8. http://www-nds.iaea.org/pgaa/egaf.html.\\
9. Neutron Cross Section, vol. 1, part A, edited by Mughabhab S. F., Divideenam M.,
 \\\hspace*{14pt}Holden N. E., Academic Press 1981.\\
10. Kadmenskij S.G., Markushev V.P., Furman W.I.,
Sov. J. Nucl. Phys. (1983) {\bf 37} 165.\\
11. Axel P.,  Phys. Rev. (1962) {\bf 126(2)} 671.\\
12. Dilg W., Schantl W., Vonach H., Uhl M., Nucl. Phys. (1973) {\bf  A217} 269.\\ 
13. Guttormsen M. at all, Nucl.\ Instrum.\
Methods Phys.\ Res.\ A (1996) \bf 374\rm  371.\\
14. Bondarenko V.A., Honzatko J., Khitrov V.A., Li Chol, Loginov Yu.E., 
\\\hspace*{14pt}Malyutenkova S.Eh., Sukhovoj A.M., Tomandl I., 
In: XII International Seminar on \\\hspace*{14pt}
Interaction of Neutrons with Nuclei, 
 Dubna, May 2004, E3-2004-169, p. 38.\\
\hspace*{14pt}http://arXiv.org/abs/nucl-ex/0406030\\
%\hspace*{14pt}http://arXiv.org/abs/nucl-ex/0410015
15. Khitrov V.A., Li Chol, Sukhovoj A.M., In: XI International Seminar on Interaction\\
\hspace*{14pt}
of Neutrons with Nuclei, Dubna, 22-25 May 2003,
E3-2004-9, Dubna, 2004, p. 98.\\
16. Sukhovoj A.M., Khitrov V.A., 
 Physics of Atomoc Nuclei (2004) {\bf  67(4)} 662\\
17. Guttormsen M. at all, Nucl.\ Instrum.\
Methods Phys.\ Res.\ A \bf 255 (1987)\rm  518.\\
18. Schiller A. et al., Nucl. Instrum. Methods Phys. Res.  (2000) {\bf A447} 498.\\
19. Sukhovoj A.M., Khitrov V.A., Li Chol,  JINR E3-2004-100, Dubna, 2004.\\\hspace*{14pt}
http://arXiv.org/abs/nucl-ex/0409016.\\
20. Khitrov V.A., Sukhovoj A.M., Pham Dinh Khang, Vuong Huu Tan,\\\hspace*{14pt} Nguyen Xuan Hai,
 In: XI International Seminar on Interaction of Neutrons with
 \\\hspace*{14pt} Nuclei,Dubna, 22-25 May 2003, E3-2004-9, Dubna, 2004, p. 107.\\
\hspace*{14pt}http://arXiv.org/abs/nucl-ex/nucl-ex/0305006.\\
21. Lone M. A., Leavitt R. A., Harrison D. A., 
 Nuclear Data Tables (1981) {\bf 26(6)} 511.\\
22. Sukhovoj A.M., Khitrov V.A.,
 Phys. Atomic Nuclei  (1999) {\bf 62} 19\\
23. Malov L.A.,
Bull. Rus. Acad. Sci. Phys. (1996) {\bf 60} 722\.\
24. Malov L.A.,  Bull. Rus. Acad. Sci. Phys. (1998) {\bf 62} 712.\\
25. Groshev L.V. et al. Atlac thermal neutron capture gamma-rays spectra,
\\\hspace*{14pt} Moscow, 1958.\\
26. Sukhovoj A.M., Khitrov V.A., Instrum. Exp. Tech.  (1984) {\bf 27} 1071.\\
27. Popov Yu.P., Sukhovoj A.M., Khitrov V.A., Yazvitsky Yu.S.,\\\hspace*{14pt}
Izv. AN SSSR, Ser. Fiz. (1984) {\bf 48} 1830.\\
28. Ignatyuk A.V., Sokolov Yu.V., Sov. J. Nucl. Phys. (1974) {\bf 19} 628.\\
\newpage
\begin{figure}%[htbp]
\vspace{-2cm}
\leavevmode
%\hspace{-.8cm}
\epsfxsize=13.0cm
\epsfbox{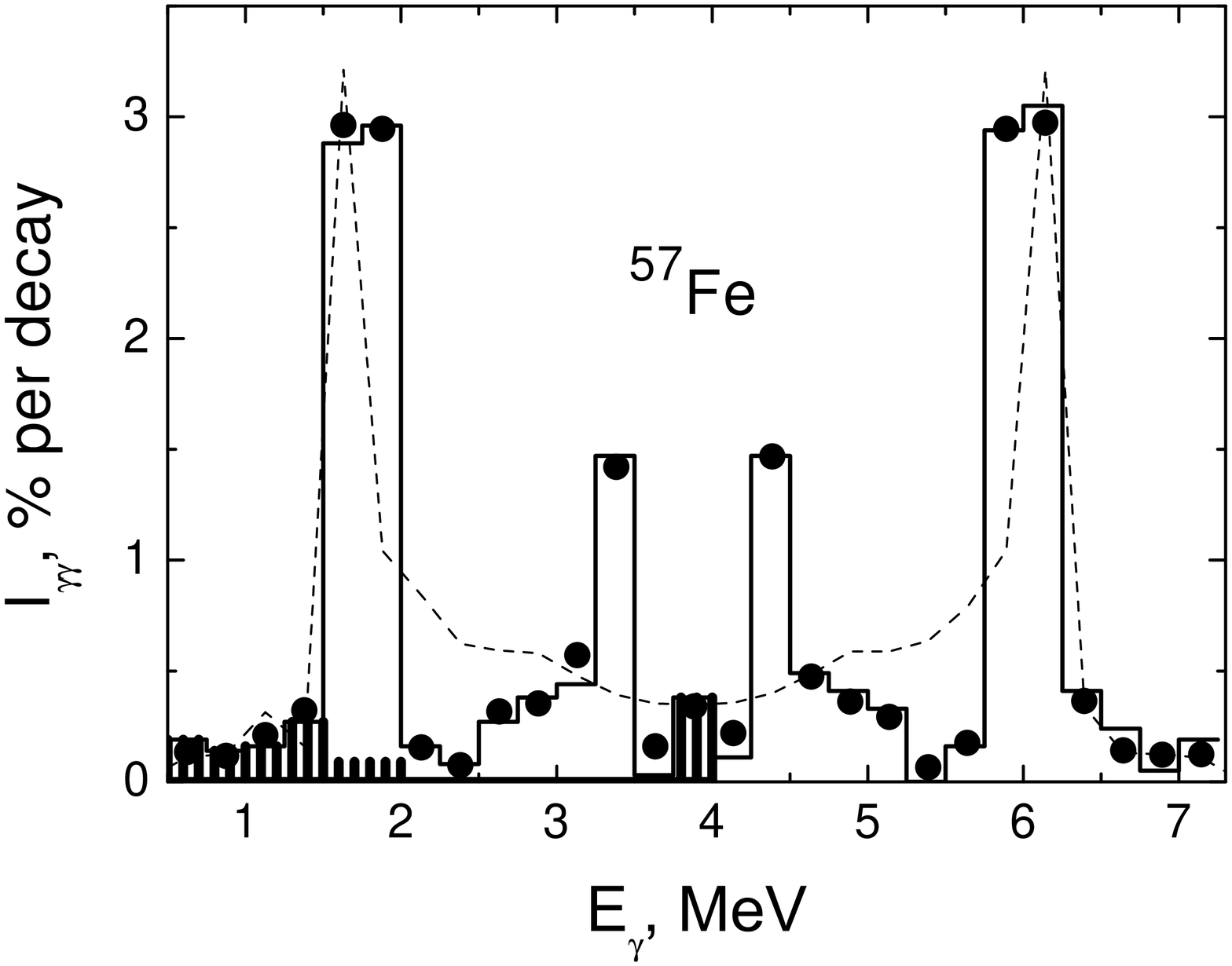}
%\vspace{1cm}

{\bf Fig. 1.} The two-step cascade intensities in $^{57}$Fe summed over the
0.25 MeV energy bins 
(histogram). Points represent an example of calculation with of
$\rho$ and $k$ obtained according to
[2]. Calculation with the data from [5,6] is shown by dashed line.
The shaded area only is used in data analyses [6].
\end{figure}

\begin{figure}%[htbp]\vspace{-3cm}
\leavevmode%\hspace{-.8cm}
\epsfxsize=13.0cm
\epsfbox{ 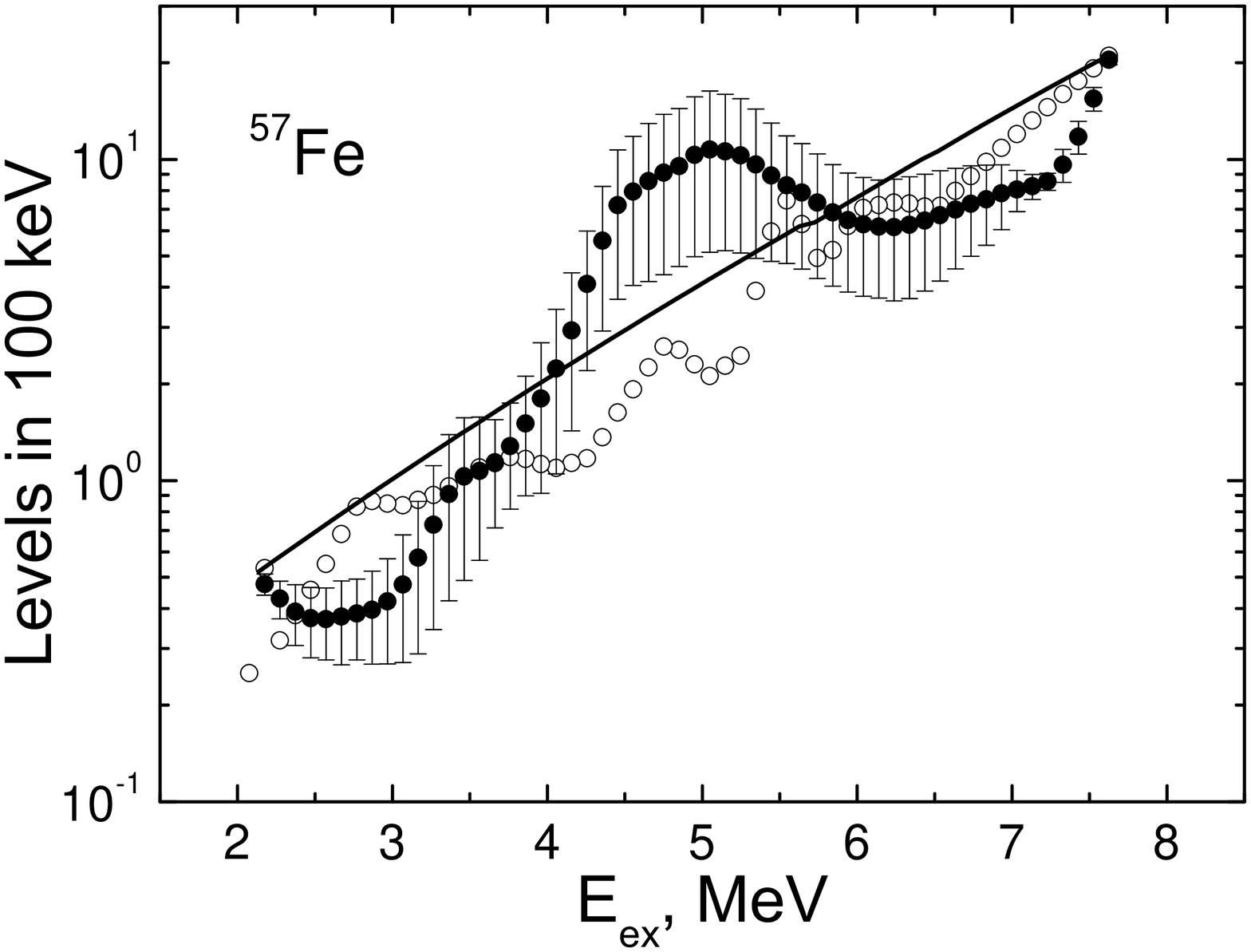}%\vspace{1cm}

{\bf Fig. 2.}
 The total number of intermediate levels of the two-step cascades reproducing 
experimental spectrum shown in Fig. 1. (points with errors). Open points show data [6] for both
parities and spins 1/2 and 3/2, line represents predictions according to model [12].
\end{figure}
\newpage
\begin{figure}%[htbp]\vspace{-2cm}
\leavevmode%\hspace{-.8cm}
\epsfxsize=13.0cm
\epsfbox{ 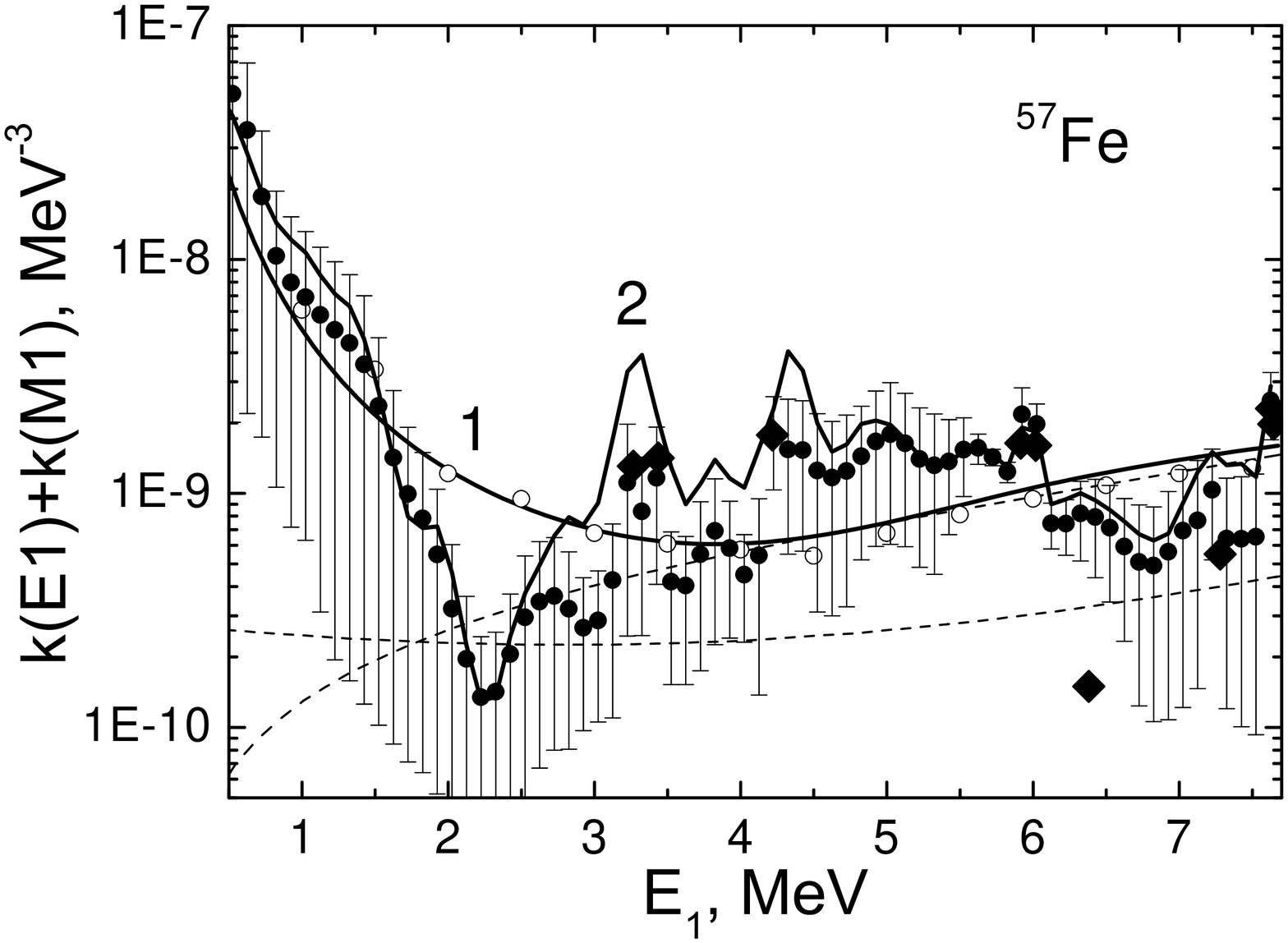}%\vspace{1cm}

{{\bf Fig. 3.}
 The possible strength functions that together with $\rho$ data from Fig. 2. reproduce
 results of [6] (points
with errors). Open points show the data [6]. Curve 1 represents $k(E1)+k(M1)$ 
used to
calculate  $I_{\gamma\gamma}$ shown as dashed line in Fig. 1. Line 2 demonstrates average
 $k$ values reproducing
together with the fixed  $\rho$ from [5] experimental cascade intensities. 
Dashed line shows predictions of k(E1) according to models [10] and [11].
The lozenges show the upper
estimation of $k$ from the maximum intensities of experimentally 
resolved primary gamma-
transitions. }\vspace{5cm}
\end{figure}

\newpage
\begin{figure}%[htbp]
\vspace{1cm}
\leavevmode%\hspace{-.8cm}
\epsfxsize=13.0cm
\epsfbox{ 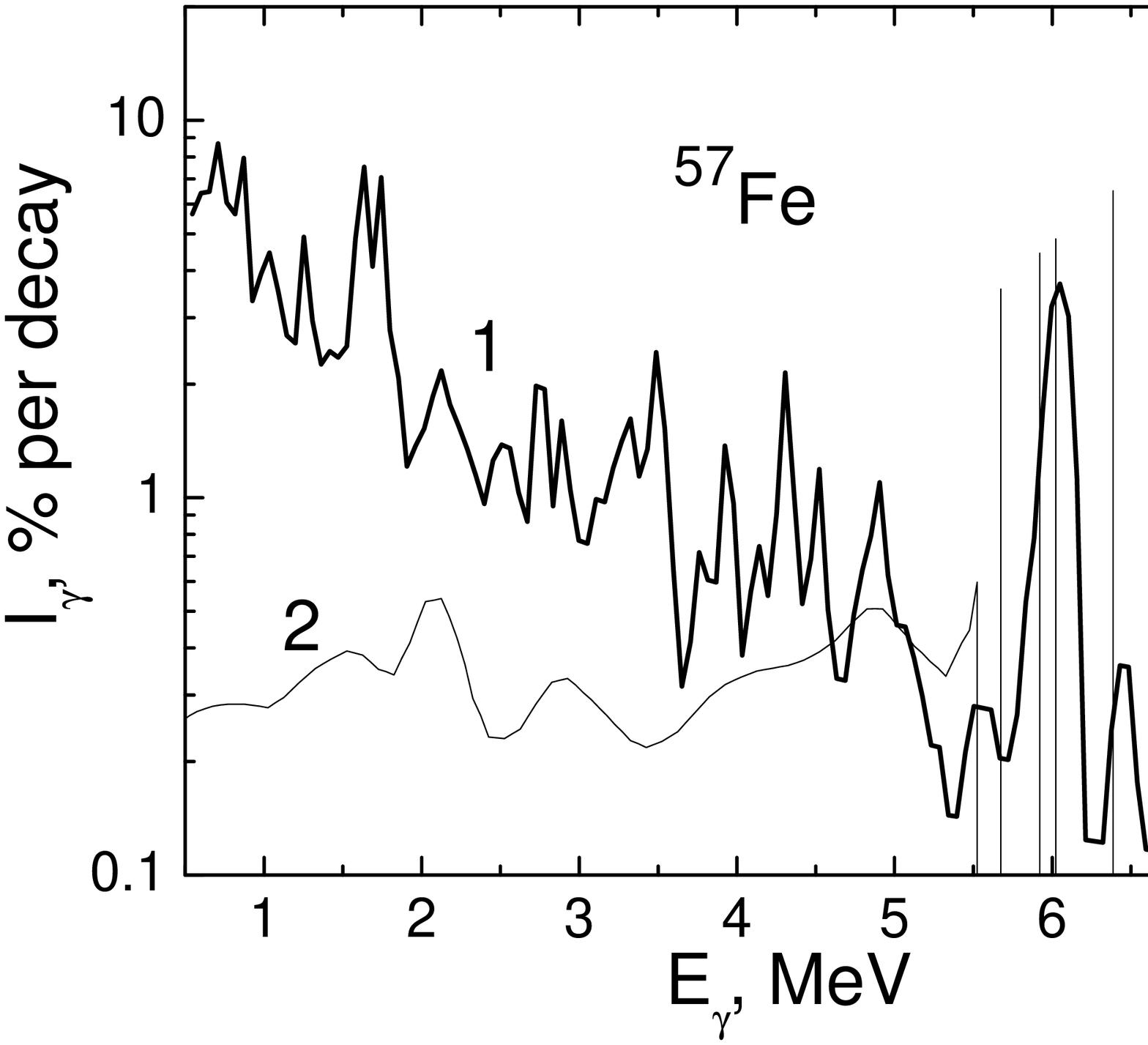}
\vspace{-3cm}

{\bf Fig. 4.} The total gamma-ray spectrum following thermal neutron radiative capture in $^{56}$Fe
(curve 1). Curve 2 represents spectrum of primary transitions for  with $\rho$ and $k$ from [5,6] (both 
spectra are normalized per 1 decay).
\end{figure}
\begin{figure}%[htbp]
\vspace{-3cm}
\leavevmode%\hspace{-.8cm}
\epsfxsize=13.0cm
\epsfbox{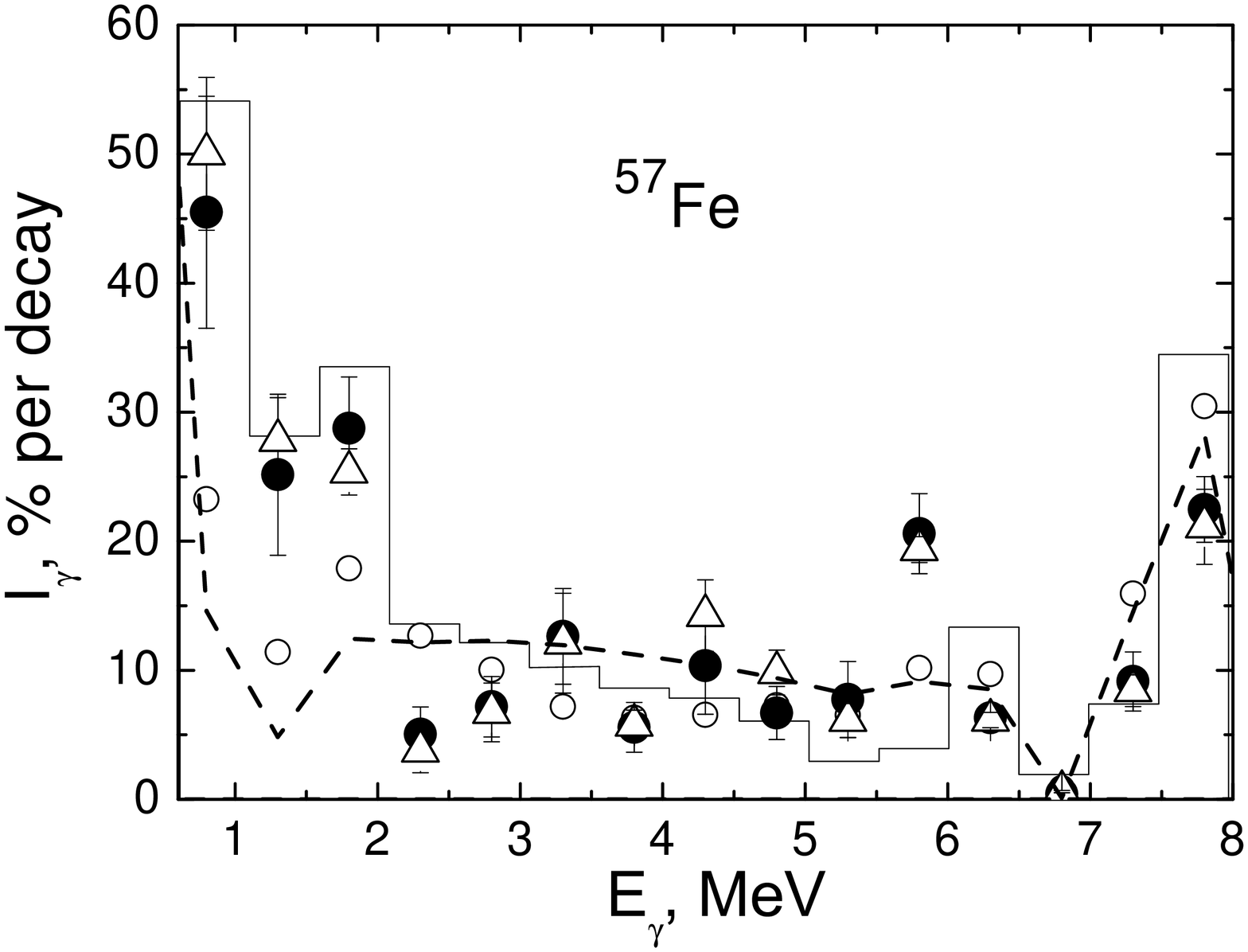}%\vspace{-1cm}\vspace{1cm}

{\bf Fig. 5.} The total gamma spectrum summed in the 1 MeV energy intervals (histogram) 
following thermal neutron capture. 
Points with errors show result of calculation with the our data given in figs. 2 and 3.
Broken dashed curve shows predictions according to models [11,12]. 
Triangles show calculation with the best [2] values of  $k$ and values of $\rho$ from [5]. 
Open circles are the calculation with $\rho$ and $k$ from [5,6] only.
 
\end{figure}

\end{document}